\documentclass[%
 reprint,
 amsmath,amssymb,
 aps,
]{revtex4-2}

\usepackage{graphicx}
\usepackage{dcolumn}
\usepackage{bm}

\usepackage{xcolor}

\begin{document}

\preprint{APS/123-QED}

\title{Stochastic Chaos and Predictability in Laboratory Earthquakes}

\author{Adriano Gualandi}
\email{adriano.gualandi@ingv.it}
\affiliation{Istituto Nazionale di Geofisica e Vulcanologia, Osservatorio Nazionale Terremoti, Rome, Italy}
\author{Davide Faranda}
\affiliation{Laboratoire des Sciences du Climat et de l'Environnement, UMR 8212 CEA-CNRS-UVSQ, Universit\'e Paris-Saclay, IPSL, 91191 Gif-sur-Yvette, France}
\affiliation{London Mathematical Laboratory, London, UK}
\affiliation{LMD/IPSL, Ecole Normale Superieure, PSL research University, Paris, France}
\author{Chris Marone}
\affiliation{La Sapienza University of Rome, Department of Earth Sciences, Rome, Italy}
\affiliation{Pennsylvania State University, Geosciences, USA}
\author{Massimo Cocco}
\affiliation{Istituto Nazionale di Geofisica e Vulcanologia, Sezione Roma 1, Rome, Italy}
\author{Gianmarco Mengaldo}
\affiliation{National University of Singapore, Department of Mechanical Engineering, College of Design and Engineering, Singapore}

\date{\today}

\begin{abstract}
Laboratory earthquakes exhibit characteristics of a low-dimensional random attractor with a dimension similar to that of natural slow earthquakes.
A model of stochastic differential equations based on rate- and state-dependent friction explains the laboratory observations.
We study the transition from stable sliding to stick-slip events and find that aperiodic behavior can be explained by small perturbations ($<1$\textperthousand) in the stress state.
Friction’s nonlinear nature amplifies small scale perturbations, reducing the predictability of the otherwise periodic macroscopic dynamics.
\end{abstract}

\maketitle

\begin{figure}
\noindent\includegraphics[width=0.5\textwidth]{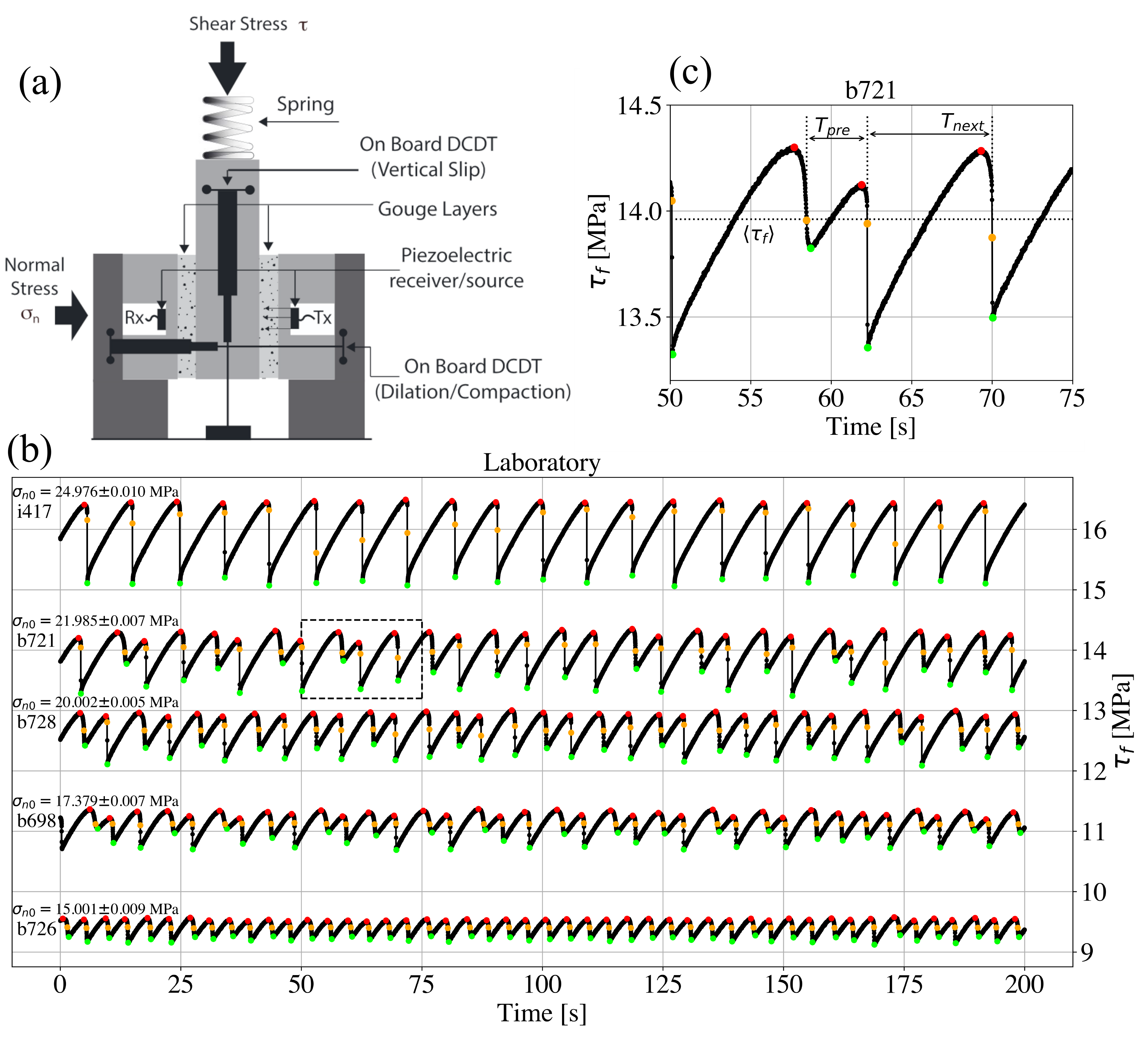}
\caption{(a) Sketch of the biaxial apparatus (BRAVA) used for the experiments \cite{Tintietal2016}. (b) Frictional shear stress time series ($\tau_f$). The system shows stick-slip behavior at different values of applied normal stress $\sigma_{n0}$. Red/green dots represent local max/min values of $\tau_f$. Orange dots are the closest points during a slip event to the average value of the shear stress $\langle \tau_f \rangle$. (c) Zoom of the black dashed box of panel (b). Interevent times are taken between the orange dots to reduce sensitivity to measurement noise associated with using max/min values and define the returning of the trajectory to a well specified region of the phase space (i.e., the hyperplane $\tau_f = \langle \tau_f \rangle$). Note that the dynamics is richer when $\frac{k}{k_c} \sim 1$ (intermediate values of $\sigma_n$), without the presence of a single characteristic labquake.}
\label{fig:fig1}
\end{figure}

Friction is a complex nonlinear phenomenon that can be modeled through many degrees of freedom (dofs) \cite{Urbakhetal2004}.
Nonetheless, laboratory studies of friction in geomaterials led to the discovery of phenomenological laws to describe the behavior of sliding surfaces with a limited number of dofs.
Rate- and state-dependent (RS) friction is today a standard framework to describe friction in a wide range of systems \cite{Dieterich1979, BaumbergerandCaroli2006, Lietal2011, Scholz2019}.
Within such formalism, the friction ($\mu$) depends on the sliding rate ($v$) and an internal state variable ($\theta$) that describes memory effects.
Stick-slip behavior, similar to that for the seismic cycle for earthquake faulting, is characterized by inter-seismic periods of load increase followed by failure events with fast or slow unloading.
Repeating failure events are explained by fault healing and changes in frictional state $\theta$ \cite{Dieterich1979, RiceandRuina1983}.
Stick-slip laboratory experiments have been used extensively to mimic the seismic cycle, for example, via a double direct shear apparatus \cite{Tintietal2016, Leemanetal2018}.
Our goal is to see if slow and fast labquakes live in the same phase space, and if they have the same number of dofs observed for slow earthquakes in nature \cite{Gualandietal2020}.
We use data from 14 stick-slip friction experiments conducted at different imposed normal stress ($\sigma_n$) conditions \cite{MeleVeeduetal2020}.
In the experiments, two layers of quartz powder are put under $\sigma_n$ and then sheared using an acrylic piston to modulate the elastic stiffness around a value of $k = 14.8$ MPa/$\mu$m \cite{Tintietal2016} (Fig.~\ref{fig:fig1}a).
The applied normal stress is used as a control parameter to systematically traverse the critical stability condition for which the stiffness of the loading apparatus ($k$) is equal to the critical stiffness of the system ($k_c \propto \frac{1}{\sigma_{n}}$) \cite{RiceandRuina1983, Guetal1984}.
We observe both slow and fast stick-slip motion, with slow/fast events occurring at low/high $\sigma_n$ values.
We deduce from the observations the number of dofs governing the system at this transition regime and set up a model with the appropriate number of phase space dimensions to match laboratory data and explain the observed range of labquake behaviors.

Each experiment includes several labquake cycles.
We use data in a $200~\textrm{s}$ time window (Fig.~\ref{fig:fig1} and Tab.~S1) in order to have a sufficient number of cycles to perform dynamical systems analysis.
We do not extend further this window to limit friction evolution effects associated with shear fabric development and wear.
Throughout each experiment, the loading velocity $v_0$ and the applied normal stress $\sigma_n$ are kept constant to some precision using servo-control.
Their mean values and standard deviations are reported in Tab.~S1.
The nominal loading velocity was $v_0 = 10$ $\mu$m/s for all experiments.
Data are sampled every $\Delta t = 0.01\textrm{ s}$.

\begin{figure}
\noindent\includegraphics[width=0.5\textwidth]{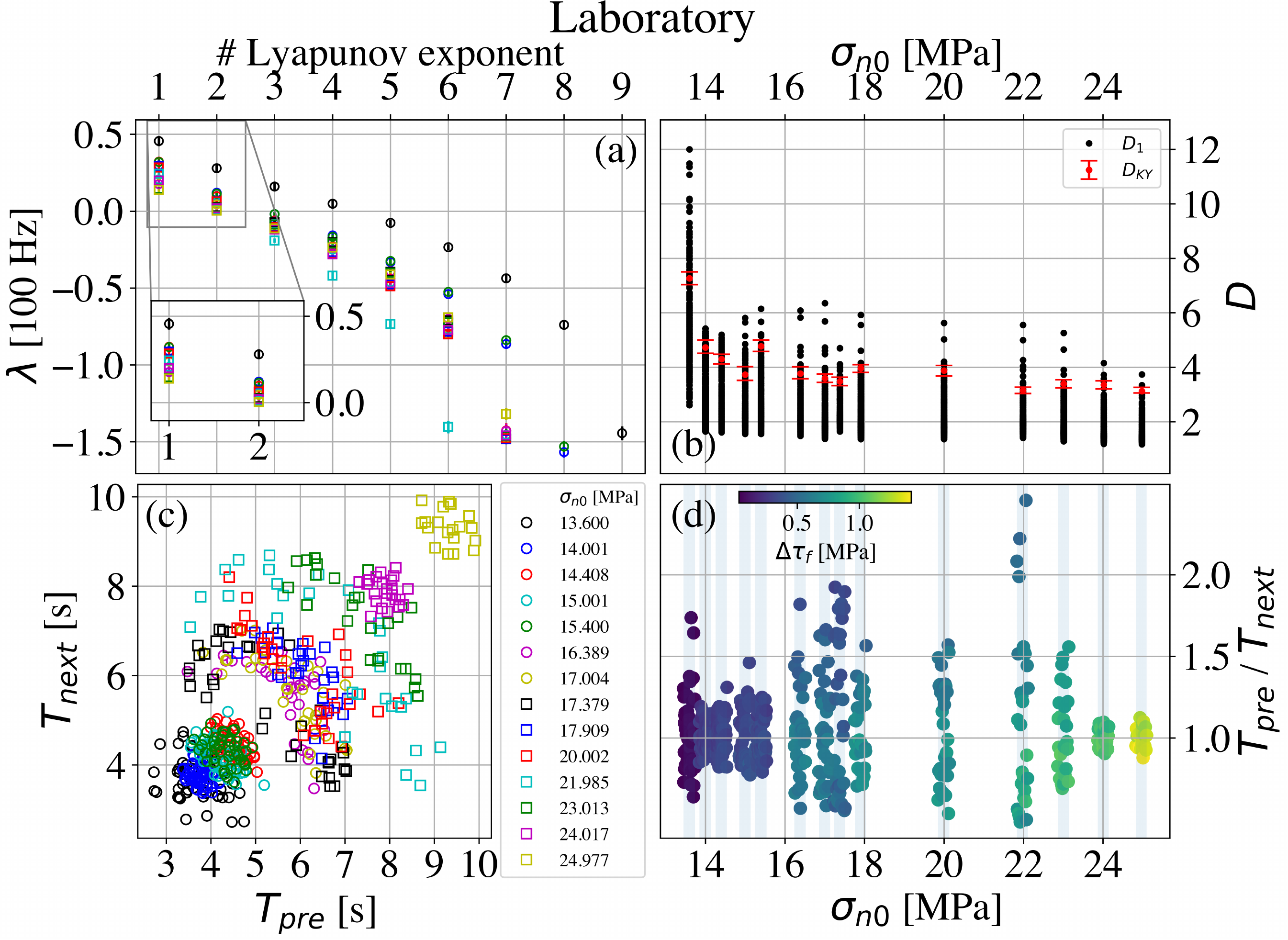}
\caption{(a) Lyapunov spectrum estimated from $\tau_f$ laboratory data. Legend is shared with panel (c). (b) Lyapunov dimension ($D_{KY})$ and information dimension ($D_1$) estimated from $\tau_f$ time series. (c) and (d): bifurcation diagrams using the interevent times of the previous ($T_{pre}$) and next ($T_{next}$) events.}
\label{fig:fig2}
\end{figure}

To infer the dimensionality of the system we use the scalar time series of the frictional shear stress ($\tau_f$, Fig.~\ref{fig:fig1}b-c) since it is the measure with the smallest noise contribution (Tab.~S2).
We calculate the Lyapunov dimension ($D_{KY}$) via the Kaplan-Yorke conjecture \cite{KaplanandYorke1979}, using the Lyapunov spectrum calculated with the method of \cite{SanoandSawada1985} (Fig.~\ref{fig:fig2}a-b).
We also estimate the information dimension ($D_1$) as the average instantaneous dimension, exploiting recent results in extreme value theory applied to dynamical systems \cite{Farandaetal2017a, Farandaetal2017b}.
With both techniques we find in most cases relatively small dimensions ($<5$) (Fig.~\ref{fig:fig2}b), suggesting that a reduced order model may suffice to explain the observations.
Details on the two approaches are provided in section S1 of the Supplementary Material.
Remarkably, similar low dimensionality was observed for slow slip events in nature \cite{Gualandietal2020}, suggesting that it might be a common feature of frictional faulting across multiple spatio-temporal scales.

We can now seek a low-dimensional model to explain the observations.
For a spring-slider model obeying RS friction, the number of variables needed to explain the dynamics (i.e., the number of dofs) is equal to 2 plus the number of internal state variables.
We notice that the dimension of the system tends to decrease with higher $\sigma_n$, i.e.\ we observe smaller dimension for faster ruptures (Fig.~\ref{fig:fig2}b).
This observation points towards the fact that inertial effects can be neglected for the conditions of our experiments.
In fact, if inertial effects were important, we would expect fast labquakes to have a higher dimension than slow events.
We thus deem reasonable to introduce a radiation damping approximation, where the inertial term is replaced by a viscous term to represent energy outflow as seismic waves \cite{Rice1993}.
This approximation reduces the number of dofs of the system by one.
For a non-inertial spring-slider system with RS friction and a single state variable it is impossible to get a bifurcation in the slip behavior: the phase space is 2-dimensional, and a closed orbit cannot split into two slipping modes without trajectory intersections occurring in the phase space.
An extra variable is needed.

\begin{figure}
\noindent\includegraphics[width=0.5\textwidth]{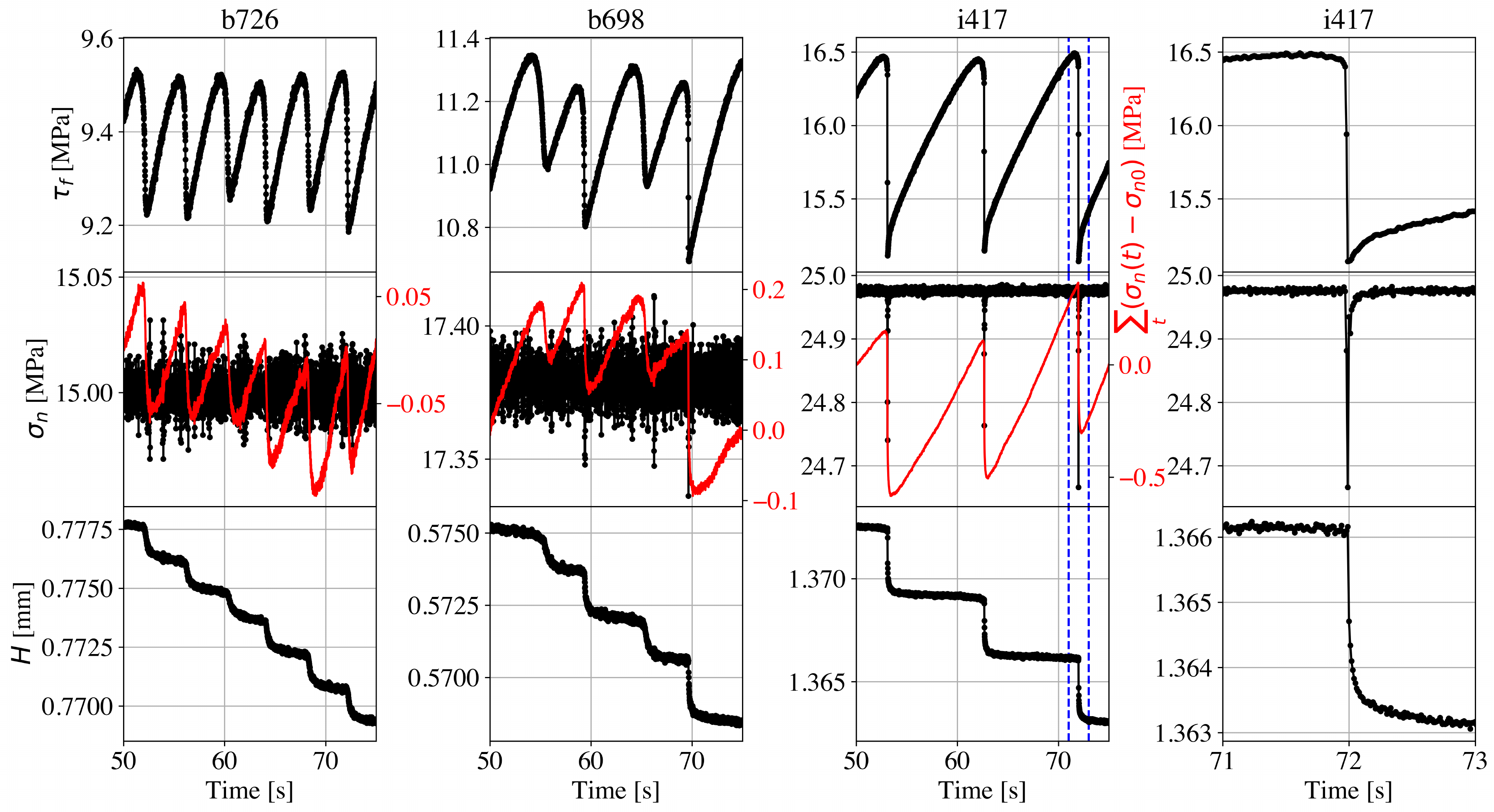}
\caption{Time evolution of shear stress ($\tau_f$, top row), normal stress ($\sigma_n$, middle row), and layer thickness ($H$, bottom row) for experiments b726 (first column), b698 (second column) and i417 (third column). Red line in the second row shows cumulative sum of the de-meaned $\sigma_n$ and clearly reveals perturbations associated with labquakes. Fourth column shows zoom (blue dashed lines in column 3) of experiment i417. Note drop in $\sigma_n$ of $\sim 0.3$ MPa and in $H$ of $\sim 2~\mu$m.}
\label{fig:fig3}
\end{figure}

In the classic spring-slider analysis, $\sigma_n$ is assumed to be constant, but experimental data shows otherwise (Fig.~\ref{fig:fig3}).
We notice a clear variation in $\sigma_n$ when $\tau_f$ drops, followed by a relaxation to recover the imposed target value of $\sigma_{n0}$.
Also the layer thickness ($H$) varies systematically over the labquake cycle, showing both a relaxation after a shear stress drop and a clear reduction (i.e., compaction) with time (Fig.~\ref{fig:fig3}).
The variations in $H$ are a function of relaxation of shear-induced dilation \cite{Maroneetal1990}, layer thinning \cite{Scottetal1994}, and the drop in $\sigma_n$ that occurs as the servo-control system responds to the labquake.
The exact partition of layer compaction between these causes is unknown.
Thus, we account for normal stress variations and dilation using the formalism of \cite{SegallandRice1995} for porous materials, where the porosity ($\phi$) and the pore pressure ($p$) are introduced as variables in the dynamics and linked to $\sigma_n$ via Terzaghi's principle \cite{Terzaghi1925}.
The only difference that we introduce is the possibility for the characteristic distance over which the porosity relaxes ($L_\phi$) to differ from the RS friction characteristic slip distance ($L$).
The model is summarized by the following set of non-dimensional ordinary differential equations (ODE) (see section S2.1 for its derivation):

\begin{subequations}
\begin{align}
    \dot{x} &=  \frac{e^x \left[ \left( \beta_1 - 1 \right) x \left( 1 + \lambda u \right) + y - u \right] + \kappa \left( \frac{v_0}{v_{{}*{}}} - e^x \right) - \dot{u} \frac{1 + \lambda y}{1+\lambda u}}{1 + \lambda u + \nu e^x} \label{eq:dx_maintext} \\
    \dot{y} &= \kappa (\frac{v_0}{v_{{}*{}}}-e^x) - \nu e^x \dot{x} \label{eq:dy_maintext} \\
    \dot{z} &= - \rho e^x (\beta_2 x + z) \label{eq:dz_maintext} \\
    \dot{u} &= - \alpha - \gamma u + \dot{z} \label{eq:du_maintext}
\end{align}
\label{eq:dx_dy_dz_du_maintext}
\end{subequations}

The state vector $\pmb{\zeta}$, i.e. that vector that fully characterizes the state of the system, is made of the non-dimensional variables $[x, y, z, u] = [\ln\left(\frac{v}{v_{{}*{}}}\right), \frac{\tau_f - \tau_0}{a\sigma_{n0}}, \frac{1}{\lambda\beta\sigma_{n0}}\left(\phi-\phi_0\right), -\frac{1}{\lambda}\frac{p}{\sigma_{n0}}]$, where $v_{{}*{}}$ is a reference sliding velocity, $\tau_0 = \mu_0 \sigma_{n0}$ is the product of the reference friction coefficient $\mu_0$ and the reference normal stress $\sigma_{n0}$, $\lambda$ is equal to $\frac{a}{\mu_0}$ with $a$ being the RS direct effect parameter (i.e., the instantaneous response of the friction coefficient to a sudden step in the sliding velocity), $\beta$ is the product of the elastic component of porosity and the combined compressibility of the fluid (air) in the pores and the elastic pores, and $\phi_0$ is a reference porosity.
The other parameters appearing in system \eqref{eq:dx_dy_dz_du_maintext} are the ratio between the RS evolutionary and direct effect parameters $\beta_1 = \frac{b}{a}$, the non-dimensional spring stiffness $\kappa$, viscous parameter $\nu$, dilatancy coefficient $\beta_2$, pore pressure in the surrounding $\alpha$, diffusivity $\gamma$, and the ratio $\rho = \frac{L}{L_\phi}$.

We adopt values for the frictional parameters derived from \cite{Tintietal2016} who used the same material as for our tests.
We adjust the values of $\beta_2, \gamma, \rho$ and $\nu$ in order to reproduce the period doubling as observed in our experiments (Fig.~\ref{fig:fig1}, section S2.2 and Tab.~S4).
For the used parameters, the solution of the ODE system does not show chaotic behavior, and the shear stress drops repeat periodically with either a single characteristic labquake or a characteristic sequence of slow and fast labquakes in a row (Fig.~S2).
In other words, with the adopted parameters the system exhibits a cyclic attractor and a limit cycle.
We notice that a quasi-static spring-slider system with two RS state variables is a particular case of the set of ODE previously derived (see section S2.3).
Such a system shows the typical behavior of the route to chaos when sufficiently reducing the stiffness of the system $\kappa$ (i.e., when increasing the reference normal stress $\sigma_{n0}$) \cite{Becker2000}.
The range of values for which deterministic chaos is observed is quite narrow ($0.06840 \lesssim \kappa \lesssim 0.06864$), and further reducing $\kappa$ leads to unstable behavior and divergence of the solution \cite{Becker2000}.
We find chaotic behavior for the system of eqs.~\eqref{eq:dx_dy_dz_du_maintext} for only a narrow range of $\sigma_{n0}$ values.
The window over which such chaotic behavior is observed depends on the selected parameters.
For the adopted parameters, such range is smaller than the minimum step in $\sigma_{n0}$, so we do not further investigate this matter here, but we do not exclude that part of the observed complexity may be due to this transition to deterministic chaos.
We note that the radiation damping term allows one to explore a larger range of values with respect to the quasi-static case, and the modifications introduced in system \eqref{eq:dx_dy_dz_du_maintext} allow us to reproduce the major feature of the data consisting in period doubling followed by a return to a single characteristic labquake cycle as $\sigma_{n0}$ increases (Fig.~S2).
To obtain these results we had to use a radiation damping term 35 times larger than what is typically adopted to simulate earthquakes.
The classically adopted value assumes an infinite fault plane in an elastic full space and radiation due only to outgoing planar shear waves \cite{Rice1993}.
These assumptions may be false in laboratory samples of few cm, and the results suggest that in the laboratory a larger fraction of energy is dissipated with respect to this idealized condition.

Period doubling in the $\tau_f$ time series  was obtained by \cite{MeleVeeduetal2020} using a boundary integral software for seismic cycle simulations.
Such model contains many more dofs with respect to the simple spring-slider model that we have implemented, and many more than those deducible from the observed time series.
Nonetheless, our model better describes the actual shape of the frictional shear stress time series (compare Figs.~S2a and S3 with Fig.~\ref{fig:fig1}b), reproduces the bifurcation at the same normal stress values actually imposed in the laboratory, and takes into account variations in normal stress (even if not perfectly, see Fig.~S4).
We thus prefer to use the low-dimensional model to describe the observations.

We further notice that, differently from the ODE simulations, the laboratory data are not exactly periodic.
This can be seen either directly from the time series of $\tau_f$, or from Fig.~\ref{fig:fig2} where we plot for each event the time to the next event ($T_{next}$) vs the time since the previous event ($T_{pre}$) (panel c) and the ratio $\frac{T_{pre}}{T_{next}}$ as a function of $\sigma_{n0}$ (panel d).
In those plots we clearly see a spread of points around the values that would represent periodic cycles of loading and failure.
Another way to measure the aperiodicity of the cycles is via the Coefficient of Variation (CV), defined as the ratio between the standard deviation and the mean of the observed interevent times.
The observed CV are reported in Fig.~S5a.
Experiments close to the transition regime (i.e., $\frac{k}{k_c} \sim 1$) show high CVs, indicating a higher complexity than experiments more distant from such a critical state.
Nonetheless, also experiments with a single characteristic labquake do not have a null CV.
We notice that $\tau_f$ and $\sigma_n$ are subject to noise fluctuations (Fig.~\ref{fig:fig3}).
These fluctuations can be described by a Wiener process $W_T$ of intensity $\varepsilon_{\tau_f}$ and $\varepsilon_{\sigma_n}$, respectively.
We estimate these intensities using the standard deviation of the time series relative to the experiment at $\sigma_{n0} = 13.6$ MPa.
In particular, we use $\varepsilon_{\tau_f} = \textrm{std}(\tau_f - \tilde{\tau}_f)$ and $\varepsilon_{\sigma_n} = \textrm{std}(\sigma_n)$, where $\tilde{\tau}_f$ is a cubic spline approximation used to smooth the time series and filter out the low frequencies.
We obtain values of $\varepsilon_{\tau_f} \simeq 0.004$ MPa and $\varepsilon_{\sigma_n} \simeq 0.006$ MPa.
Adding only observational noise to the ODE simulations does not explain the observed CV variability (Fig.~S6).
Introducing such noises in terms of $y$ and $u$ leads to a set of stochastic differential equations (SDE) with eqs.~\eqref{eq:dy_maintext} and \eqref{eq:du_maintext} being modified into (section S2.1):

\begin{subequations}
\begin{align}
    \textrm{d}y &= \left[ \kappa \left(\frac{v_0}{v_{{}*{}}}-e^x \right) - \nu e^x \dot{x} \right] \textrm{d}T + \varepsilon_y \textrm{d}W_T \label{eq:dy_W_maintext} \\
    \textrm{d}u &= \left[ - \alpha - \gamma u + \dot{z} \right] \textrm{d}T + \varepsilon_u \textrm{d}W_T \label{eq:du_W_maintext}
\end{align}
\label{eq:dy_du_W_maintext}
\end{subequations}

\begin{figure}
\noindent\includegraphics[width=0.5\textwidth]{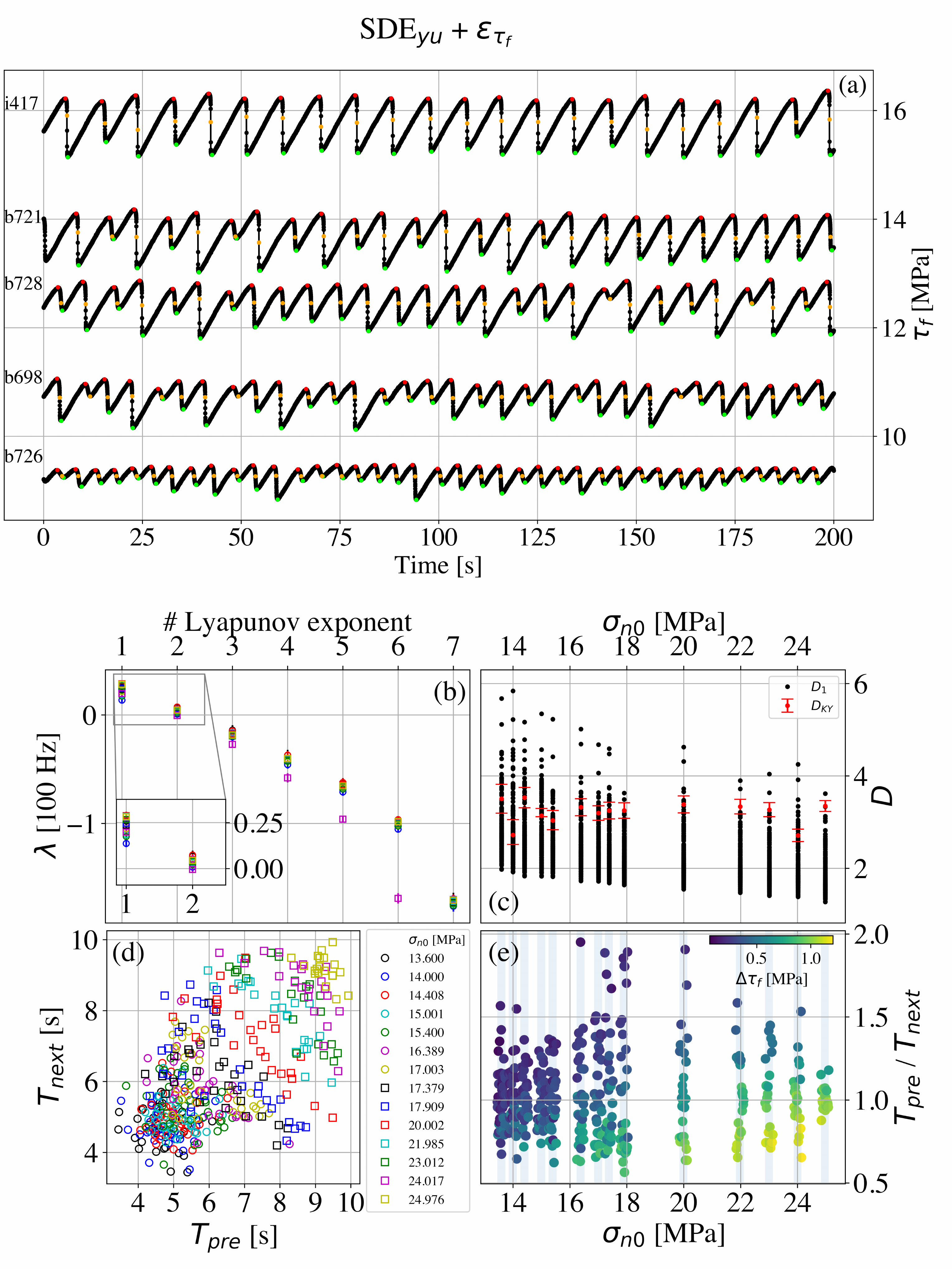}
\caption{(a) Frictional shear stress $\tau_f$ simulated using the system of SDE with stochastic terms added to $y$ and $u$, and addition of observational noise $\varepsilon_{\tau_f}$ on $\tau_f$. Imposed $\sigma_{n0}$ to mimic experiments b726, b698, b728, b721 and i417. (b)-(e): As Fig.~\ref{fig:fig2}, but for SDE simulated time series with stochastic terms added to $y$ and $u$ and the addition of observational noise $\varepsilon_{\tau_f}$ to $\tau_f$.}
\label{fig:fig4}
\end{figure}

The simulated time series are shown in Fig.~S7.
Of course, also $v$ and $\phi$ (i.e., $x$ and $z$) may be subject to noise that can perturb the dynamics of the system.
Contrary to $\tau_f$ and $\sigma_n$, we do not directly measure them.
Before introducing further complications, we consider the noises on $\sigma_n$ and $\tau_f$ as the most relevant ones.
To mimic the laboratory time series, we finally add measurement noise of $\varepsilon_{\tau_f}$ to the SDE generated $\tau_f$ time series (Fig.~\ref{fig:fig4}a).
The Lyapunov spectrum, the system dimension, and the interevent times are shown in Figs.~\ref{fig:fig4}b-e.
We find a Lyapunov spectrum similar to the one derived from the data, $D_{KY}$ between 3 and 4, and both $D_{KY}$ and $D_1$ decreasing with increasing $\sigma_{n0}$ (see Fig.~\ref{fig:fig2}b for comparison).
Furthermore, in the bifurcation diagrams we see a spread similar to the one observed in the laboratory time series.
This result suggests that the aperiodicities in the returning time of laboratory earthquakes (of any type, either slow or fast or mixed) could be the result of a stochastic noise component that enters the dynamics and gets amplified due to the nonlinearities of the equations.
As a consequence, with the current available set up, stick-slip cycles near the critical transition regime obtained from laboratory experiments can be reproducible only statistically and not deterministically, in the sense that, with the same rocks and conditions applied, we can expect to reproduce similar average interevent times and standard deviations, but not the same exact sequence of ruptures.
Similar observations were made for a different frictional system of labquakes \cite{KarnerandMarone2000}, and this type of feedback between the apparatus' vibrations and the system dynamics has been documented for turbulent flows \cite{Farandaetal2017a}.
These behaviors may be a more general characteristic of nonlinear frictional systems.

\begin{figure*}
\noindent\includegraphics[width=\textwidth]{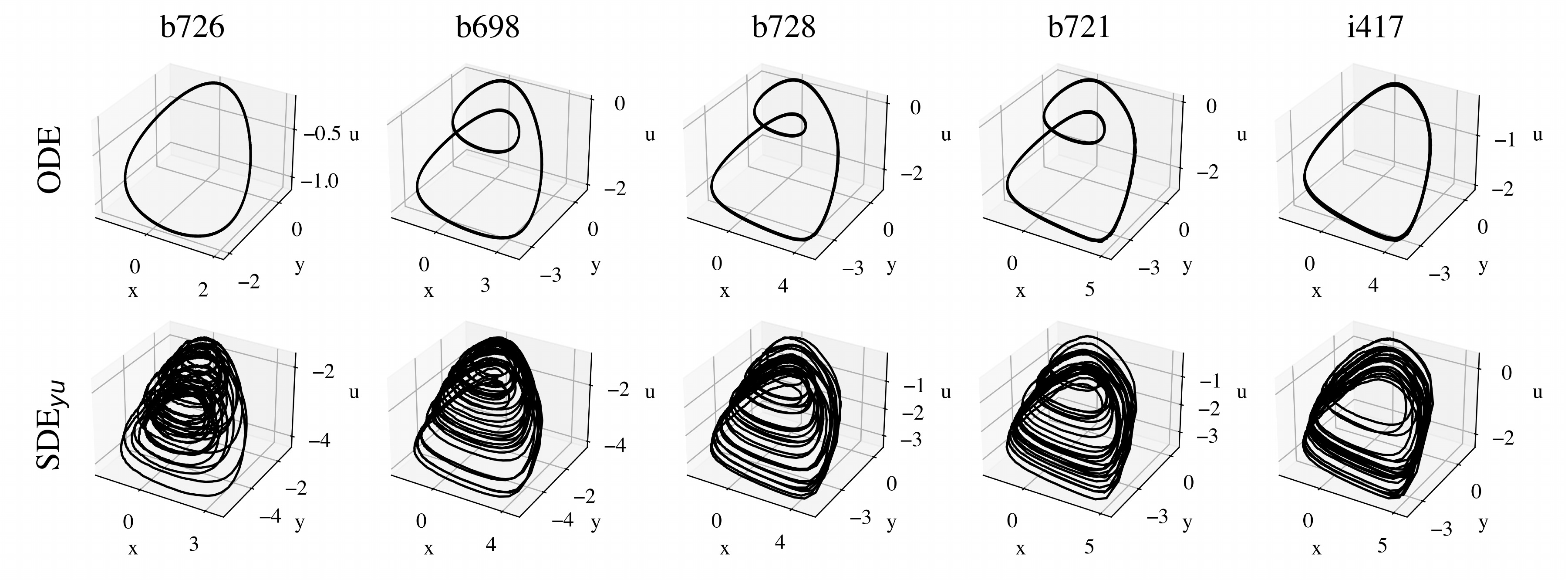}
\caption{Phase space subregion $xyu$ of the ODE (top) and SDE (bottom) simulations for the experiments b726, b698, b728, b721 and i417.}
\label{fig:fig5}
\end{figure*}

Our results point towards the existence of a random attractor (Fig.~\ref{fig:fig5}) to describe the seismic cycle: minimal variations of the order of less than 1\textperthousand\ on the shear and normal stresses applied to the fault (intended as the experimental frictional interface) influence the large scale dynamics and the recurrence time of a rupture, inducing CV of a few percent points ($> 3\%$, Fig.~S5a).
The implications on natural faults are quite relevant.
At seismogenic depths ($\sim10$ km), variations of less than 1\textperthousand\ of the lithostatic stress would correspond to hundreds of kPa or less.
Possible causes that can generate stress perturbations of 10-100 kPa are: other tectonic sources (with both static and dynamic stress variations \cite{Freed2005}), magmatic intrusions in volcano-tectonic environments \cite{Chenetal2019}, surface atmospheric loading \cite{DAgostinoetal2018, Pintorietal2021}, anthropogenic activity, and solid tides \cite{Rubinsteinetal2008}.
The laboratory experiments here considered were conducted to explore the transition regime $\frac{k}{k_c}\sim1$, which corresponds to regions of transition between linearly stable to unstable behavior.
Such a transition zone might be similar to the one where episodic tremors are observed in nature, since we found that labquakes at the transition regime have dimension similar to the one of slow earthquakes in Cascadia \cite{Gualandietal2020}.
Given that $k_c$ depends on $\sigma_n$ \cite{RiceandRuina1983}, variations $\varepsilon_{\sigma_n}$ can influence the stability of the system, making it sensitive to small stress perturbations.
This suggests that we should not treat faults as isolated systems, especially when modeling the limits of the seismogenic zone in nature.

The lack of an appropriate deterministic description of all the aforementioned stress perturbations may explain why earthquake forecast is a difficult task and statistical methods are used \cite{Main1996, Main1999}.
Nonetheless, the addition of the available physical information is a fundamental aspect to improve our forecasts.
For example, a commonly used statistical distribution to describe the recurrence of a characteristic earthquake is the one derived from the study of a Brownian Relaxation Oscillator (BRO).
This is a time and slip predictable stick-slip system subject to a stochastic perturbation of its ``load state".
The distribution that describes the expected waiting time of the next event, given the (physical) knowledge of the current ``load state" of the system, is the so-called Brownian Passage Time (BPT) distribution \cite{Matthewsetal2002}.
Our results relying on laboratory experiments may explain why a mix of physical knowledge and statistical methods like those obtained with a BPT model are often used for earthquakes forecast \cite{Ogata2017}.
Compared to the BRO, here we have introduced the stochastic perturbations under a more rigorous physical description of friction, i.e.\ under the RS framework using two length-scale parameters.
Despite a low average dimension of the attractor, the maximum instantaneous dimension deduced from the data ranges from $\sim12$ (for experiment b727, $\sigma_{n0} = 24.017$ MPa) to $\sim47$ (for experiment b695, $\sigma_{n0} = 17.909$ MPa).
This suggests that extra dofs are needed to fully characterize the dynamics in some regions of the phase space.
Supplying a stochastic term to the dynamics is an admittance of our ignorance on how to monitor and describe these extra dofs \cite{Vere-Jonesetal2005}.
The maximum instantaneous dimension increases from $\sim2$ to $\sim38$ when perturbing the dynamics with a stochastic term, and it goes up to $\sim83$ when introducing also observational noise.
The model we propose is one of the possible candidates and it is based on the modelling of small scale fluctuations via stochastic terms.
This choice is not unique and deterministic descriptions could hold as well.
Whether solids or fluids should be phenomenologically modelled via stochastic or deterministic equations is still an open problem \cite{Nathetal2009, Cruzeiro2020} which affects, for example, the quality of weather forecasts and climate predictions \cite{Palmer2019}.
Furthermore, another limitation may come from the fact that instabilities are introduced by inertial effects, that certainly play a significant role for natural earthquakes.
Despite these limitations, the proposed description constitutes a step forward towards a model that can explain major features of the laboratory shear stress time series and reconcile the number of dofs that can be deduced from the observations.
Furthermore, this model can explain fluctuations of the labquakes stress drop caused by the inclusion of stochastic perturbations to the dynamics, thus contributing to the debate on stress drop scaling with earthquake size \cite{Coccoetal2016}.
Thanks to the ODE and SDE models, we can estimate the effect of the stochastic term to the predictability of the dynamics. We calculate the Lyapunov time ($t_{Lyap}$) using the maximum Lyapunov exponent of the spectrum obtained from clean time series, i.e. without additional observational noise (Figs.~S5b-c).
The limited $t_{Lyap}$ for ODE models is due to the finite $\Delta t$.
For SDE models $t_{Lyap}<1$ s.
While there exist model-free techniques capable of predicting the behavior of chaotic systems up to $8t_{Lyap}$ into the future \cite{Pathaketal2018}, as future steps we envision the application of data assimilation approaches that can use the model here presented to advance the state of the system.
Similar to weather forecast, we think that ensemble forecast is the most reasonable way to assess the future state of the system because of the unavoidable stochastic terms that affect the dynamics.
We do not exclude that this model may also be used to improve laboratory earthquakes forecast with machine learning approaches, generating at will synthetic data that resembles the major characteristics of the observed time series and reducing the shortcoming of limited training datasets.

\acknowledgments
We thank Carolina Giorgetti, Luca Dal Zilio, Giacomo Pozzi, Lauro Chiaraluce and Marco Maria Scuderi for insightful discussions.

\bibliography{biblio.bib}

\end{document}